\documentclass[pra,twocolumn,showpacs]{revtex4}
\usepackage{amsfonts}
\usepackage{amsmath}
\usepackage{amssymb}
\usepackage{color}
\usepackage{subfigure}
\usepackage{graphicx} 
\usepackage[varg]{txfonts}

\newcommand{\zetaa}[2]{\zeta_{#1}(#2)}
\newcommand{\eqn}[1]{\begin{eqnarray}#1\end{eqnarray}}
\newcommand{\x}{\mathbf{x}}
\newcommand{\etal}{\emph{et al.}}

\newcommand{\mbf}[1]{\mathbf{#1}}

\newcommand{\sref}[1]{Sec.~\ref{#1}}
\newcommand{\eref}[1]{Eq.~(\ref{#1})}
\newcommand{\eeref}[1]{(\ref{#1})}
\newcommand{\fref}[1]{Fig.~\ref{#1}}
\newcommand{\NOTE}[1]{}
\newcommand{\NOTES}[1]{}
\def\NOTES{
\setlength{\marginparwidth}{3.5cm}
\setlength{\evensidemargin}{3.5cm}
\renewcommand{\NOTE}[1]{\marginpar{\sf\small \color{red} ##1}}}
\newcommand{\DRAFT}[1]{}
\def\DRAFT{
\setlength{\evensidemargin}{7cm}
\usepackage{showlabels}
\NOTES
}

\begin{document}
\title{Scale invariant thermodynamics of a toroidally trapped Bose gas}
\author{A. S. Bradley} 
\affiliation{Jack Dodd Centre for Quantum Technology, Department of Physics, University of Otago, Dunedin, New Zealand}
\date{\today}
\begin{abstract} 
We consider a system of bosonic atoms in an axially symmetric harmonic trap augmented with a two dimensional repulsive Gaussian optical potential. We find an expression for the grand free energy of the system for configurations ranging from the harmonic trap to the toroidal regime. For large tori we identify an accessible regime where the ideal gas thermodynamics of the system are found to be independent of toroidal radius. This property is a consequence of an invariant extensive volume of the system that we identify analytically in the regime where the toroidal potential is radially harmonic. In considering corrections to the scale invariant transition temperature, we find that the first order interaction shift is the dominant effect in the thermodynamic limit, and is also scale invariant. We also consider adiabatic loading from the harmonic to toroidal trap configuration, which we show to have only a small effect on the condensate fraction of the ideal gas, indicating that loading into the scale invariant regime may be experimentally practical. 
\end{abstract}
\maketitle

\section{Introduction}
While Bose-Einstein condensation of dilute gases is now routinely observed, the degree of control and interrogation now affords more detailed studies of the dynamics of the condensation process, such as the symmetry breaking associated with the Kibble-Zurek mechanism (KZM)~\cite{Kibble1976,Zurek1985,Anglin1999,Dziarmaga2008a}. 
Recent experiments loading toroidal traps with Bose-degenerate gases~\cite{Ryu07a,Weiler08a} have shown the need for a basic theoretical understanding of the properties of Bose-Einstein condensation in toroidal traps, which may be important for future studies of BEC formation and tests of KZM. 

Following the development of storage rings for neutral atoms~\cite{Sauer01a}, toroidal traps for BECs were proposed using various combinations of magnetic and optical techniques~\cite{Wright00a,Arnold04a}. Sagnac interferometry is an important application for large toroidal traps, having the feature of resolution proportional to the area of the interferometer~\cite{Gustavson00a}. While traps of order $\sim 3$ mm diameter have been created~\cite{Gupta2005,Arnold06a}, BEC loading for such large traps was limited to launching the BEC into the toroid which then acts as a dispersive waveguide. Smaller toroids, which can be more easily loaded, have recently been produced~\cite{Ryu07a,Schnelle08a,Weiler08a,Heathcote08a} opening the way for studies of superfluidity and persistent currents in a nontrivial trapping topology. 

Theoretical efforts have focussed on BECs far below $T_c$. Many features of toroidal geometry have been studied, including topological phases~\cite{Petrosyan99a}, the stability of macroscopic persistent currents~\cite{Javanainen98a,Benakli99a,Salasnich99a,Jackson06a,Ogren07,Ogren09} , excitation spectra~\cite{Nugent03a}, atomic phase interference devices~\cite{Anderson03a}, vortex-vortex interactions~\cite{Schulte02a}, generation of excitations via stirring~\cite{Brand01a}, dynamics of sonic horizons~\cite{Jain07a}, parametric amplification of phonons~\cite{Modugno06a}, rotational current generation~\cite{Bhattacherjee04a}, the interplay of interactions and rotation~\cite{Kavoulakis04a}, giant vortices~\cite{Cozzini05a}, and vortex signatures~\cite{Cozzini06a}. Ideal gas theory has recently been used~\cite{Kling08a} to study the rapidly rotating Bose gas in a quartically stabilized harmonic trap realized at ENS~\cite{Bretin04a}. The BEC transition temperature in non-power law traps is thus becoming more relevant, and a recent study of optical lattices~\cite{Blakie07a} gives further indication that analytical expressions can be found for increasingly rich potentials. However, the ideal gas thermodynamics of three dimensional toroidal potentials--crucial for understanding the dynamics of Bose-Einstein condensation--have not been addressed. 

The effect of trapping geometry on the BEC transition was emphasized by Bagnato \etal~\cite{Bagnato87a}. It was observed that, for power law traps, increasing the confinement of the system has the effect of increasing the peak phase space density and thus raising the temperature of the BEC transition. An understanding of how phase space density depends on the toroid size is crucial for making large ultra-cold persistent currents. There is  also the role of topology to consider. In particular, in a toroidal trap the angular spatial coordinate becomes unavailable for thermalization, suggesting a potentially interesting interplay between topology and system size.

In this work we study the properties of a Bose gas trapped by a specific toroidal potential. The potential is created from a harmonic magnetic potential combined with a repulsive (blue detuned) optical potential with a Gaussian spatial profile~\cite{Weiler08a}; we refer to this potential as harmonic-Gaussian and show that it has uniquely interesting properties which advantage it for creating large toroidally trapped BECs. Using the semiclassical approach to the thermodynamics of the ideal gas we find an exact expression for the free energy. 

Examining the properties of the system for increasing toroidal radius shows the existence of an analytically tractable regime of \emph{scale invariance} with respect to the toroid radius. In this regime the toroidal trap is well approximated as radially harmonic. We use the theory of Romero-Rochin~\cite{RomeroRochin05a} to identify the invariant generalized extensive volume of the system which governs scale invariance. 
We further generalize this result to show that the system enters a scale invariant regime even when this approximation is not valid. Focusing on the preservation of quantum degeneracy, we then treat the BEC transition temperature in detail and consider finite size and mean field corrections to the scale invariant result. Finally, we discuss possible means to reach the scale invariant regime.
\subsection{Geometry of the harmonic-Gaussian potential}
We consider a Bose gas confined in the trapping potential
\begin{equation}\label{Vdef}
V(\textbf{x})=\frac{m}{2}\left(\omega_r^2r^2+\omega_z^2z^2\right)+V_{\rm OP}(r),
\end{equation}
where axial and radial trapping frequencies are $\omega_z$, $\omega_r$, and the non-harmonic potential is given by
\begin{equation}\label{OPdef}
V_{\rm OP}(r)=V_{0}\exp{(-r^2/\sigma_{0}^2)},
\end{equation}
where $r$ is the distance from the $z$-axis. This potential can be created using a magnetic trap combined with a detuned laser field propagating along the $z$-axis which forms an optical dipole potential~\cite{Grimm01a}. 
\begin{figure}[t]
\includegraphics[width=\columnwidth]{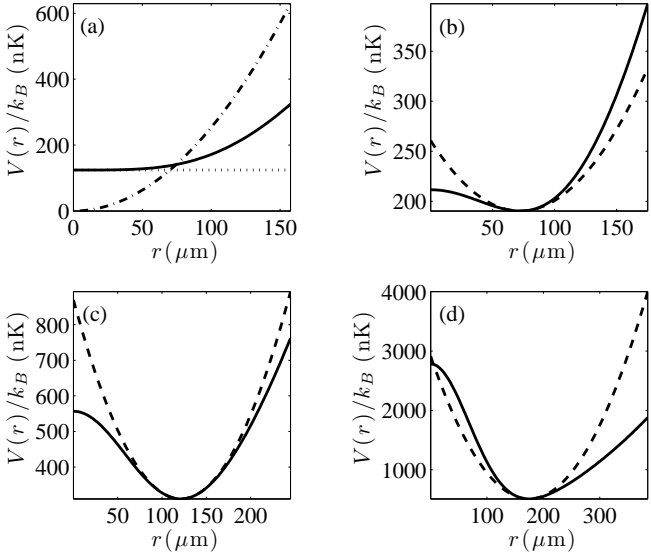}
\caption{The harmomic-Gaussian trapping potential is shown at $z=0$ (solid lines) for an optical potential of width $\sigma_0=70\mu$m and (a) $V_0/k_B=124\;{\rm nK}$, (b) $V_0/k_B=211\;{\rm nK}$, (c) $V_0/k_B=557\; {\rm nK}$, and (d) $V_0/k_B=2784 \; {\rm nK}$. In (a) the dashed-dotted curve is the bare harmonic potential $(V_0=0)$ and the dotted line is $V_\sigma\equiv m\omega_r^2\sigma_0^2/2$, which in this case is also the critical case for bifurcation of the trap minimum: $V_0= V_\sigma$. The dashed lines in (b)--(d) give the harmonic approximation to the radial potential. The harmonic trap frequencies are $(\omega_z,\omega_r)=2\pi (15.3,7.8)$ Hz}
\label{fig1}
\end{figure}
To determine the geometry of the trap it is convenient to define the energy 
\eqn{\label{Vs}
V_\sigma\equiv \frac{1}{2}m\omega_r^2\sigma_0^2,
}
which is the potential energy of an atom at $r=\sigma_0$ in the harmonic potential; this energy will play a central role in determining the thermodynamics of the system.
The minimum of the combined potential is located at $z=0$ and $r=r_m$, where
\begin{equation}
\begin{split}
r_m\equiv
\begin{cases} 
0 &\text{for \hspace{2mm}$V_0< V_\sigma$,}
\\
\sqrt{\sigma_0^2\ln{(V_0/V_\sigma)}} & \text{for \hspace{2mm}$V_0\geq V_\sigma$,}
\end{cases}
\end{split}
\end{equation}
from which we find that $V_m\equiv {\rm min}[V(\x)]=V(r=r_m,z=0)$ is given by
\begin{equation}\label{Vmdef}
\begin{split}
V_m\equiv
\begin{cases} 
V_0 &\text{for \hspace{2mm}$V_0< V_\sigma$,}
\\
V_\sigma(1+\ln{(V_0/V_\sigma)})& \text{for \hspace{2mm}$V_0\geq V_\sigma$.}
\end{cases}
\end{split}
\end{equation}
There are three distinct regimes parametrized by the ratio $V_0/V_\sigma$ which are
\begin{enumerate}
\item \emph{Dimple trap} ($V_0<0$) The Gaussian forms a dimple in the center of the harmonic trap.
\item \emph{Flat trap} ($0<V_0\leq V_\sigma$) The trap is flattened but not toroidal, as shown in \fref{fig1} (a).
\item \emph{Toroidal trap} ($V_\sigma<V_0$) The trap becomes toroidal, as shown in \fref{fig1} (b)-(d). 
\end{enumerate}

\section{Ideal Bose gas in a harmonic-Gaussian trap}
In this section we develop the general grand-canonical theory of the ideal Bose gas in the harmonic-Gaussian trap.

\subsection{Grand canonical free energy}
In general, the grand potential function for the system is
\begin{equation}
{\cal F} = -k_B T\ln {\cal Z},
\end{equation}
where ${\cal Z}$ is the grand partition function. For the Bose gas distributed over levels with excitation energy $\epsilon_i$ this becomes
\begin{equation}
{\cal F}  = k_B T\sum_i \ln\left(1-e^{\beta(\mu-\epsilon_i)}\right),
\end{equation}
 $\beta=1/k_BT$.
For the situation of interest, the chemical potential is assumed to approach the ground state energy of the trap $\epsilon_0$, leading to macroscopic occupation of the ground state $N_0$. In this regime, using the semi-classical approximation for the excited states, ${\cal F} $ can be written as 
\begin{equation}
{\cal F} =N_0(\epsilon_0-\mu)-\left(\frac{m}{2\pi\hbar^2}\right)^{3/2}\sum_{k=1}^\infty\frac{e^{k\beta\mu}}{(k\beta)^{5/2}}{\cal G}(k\beta),
\label{freeE}
\end{equation}
where
\begin{equation}\label{Gdef}
{\cal G}(\alpha)\equiv\int d^3\mbf{x}\; \exp{\left[-\alpha V(\mbf{x})\right]}.
\end{equation}
All thermodynamic properties are determined by ${\cal G}(\alpha)$ and its derivatives. The BEC transition temperature is found from the total atom number at the point where the chemical potential reaches the ground state energy
\begin{equation}
N=-\frac{\partial {\cal F}}{\partial \mu}\Bigg{|}_{\mu=\epsilon_0},
\end{equation}
with $N_0=0$, giving the transition temperature $T_c$ as the solution of 
\begin{equation}
N \lambda_{dB}^3=\sum_{k=1}^\infty \frac{e^{k\beta V_m}}{k^{3/2}}{\cal G}(k\beta),
\end{equation}
where $\lambda_{dB}\equiv\sqrt{2\pi\hbar^2/mk_BT}$ is the thermal de Broglie wave-length.
\subsection{Thermodynamics of the harmonic-Gaussian trap}
For the general form of the harmonic-Gaussian trap \eeref{Vdef} we can evaluate \eeref{Gdef} to find 
\begin{equation}
{\cal G}(\alpha)=2\pi\sigma_0^2\sqrt{\frac{\pi}{\alpha 2m\omega_z^2}}\frac{\gamma(\alpha V_\sigma,\alpha V_0)}{(\alpha V_0)^{\alpha V_\sigma}},
\end{equation}
where
\begin{equation}
\gamma(a,x) = \int_0^x e^{-t} t^{a-1} dt.
\end{equation}
is the incomplete Gamma function. 

 In the semiclassical approximation it is consistent to take the ground state energy as the trap minimum $\epsilon_0\to V_m$, and the free energy can now be written as
\begin{equation}
{\cal F}=N_0(V_m-\mu)-\frac{\zeta_4(e^{\beta \mu},\beta V_\sigma,\beta V_0)}{\beta^4\hbar^3\omega_r^2\omega_z},
\end{equation}
where we define the generalized $\zeta$-function 
\begin{equation}\label{zetaDef}
\zeta_\nu(z,a,b)\equiv a\sum_{k=1}^\infty \frac{z^k}{k^{\nu-1}}\Gamma(ka)\gamma^*(ka,kb),
\end{equation}
and $\gamma^*(a,x)=x^{-a}\gamma(a,x)/\Gamma(a)$ is a single valued analytic function of $a$ and $x$ with no singularities~\cite{Abram}. This function has the following asymptotics:
\begin{eqnarray}
\lim_{V_0\to 0}\zeta_\nu(e^{\beta \mu},\beta V_\sigma,\beta V_0)&=&\zeta_\nu(e^{\beta\mu}),\label{Vzero}\\
\lim_{\frac{V_0}{V_\sigma}\to \infty;\;\frac{V_\sigma}{k_BT}\to \infty}\frac{\zeta_\nu(e^{\beta \mu},\beta V_\sigma,\beta V_0)}{\sqrt{2\pi \beta V_\sigma}}&=&\zeta_{\nu-1/2}(e^{\beta(\mu-V_m)}),\label{Vinf}\\
\lim_{V_0\to -\infty}\zeta_\nu(e^{\beta \mu},\beta V_\sigma,\beta V_0)\frac{|V_0|}{V_\sigma}&=&\zeta_\nu(e^{\beta(\mu+|V_0|)}),\label{Vninf}
\end{eqnarray}
where $\zeta_\nu(z)=\sum_{k=0}^\infty z^k/k^\nu$ is the polylogarithm function. In the regime of Bose-Einstein condensation, where $\mu- V_m\to0$ this further reduces to the ordinary Riemann-zeta function $\lim_{V_0\to 0}\zeta_\nu(1,\beta V_\sigma,\beta V_0)=\zeta_\nu(1)=\zeta(\nu)$. 

The number of atoms in the system is given by
\begin{equation}\label{eq:N}
N=-\frac{\partial {\cal F}}{\partial \mu}=N_0+\frac{\zeta_3(e^{\beta \mu},\beta V_\sigma,\beta V_0)}{\beta^3\hbar^3\omega_r^2\omega_z},
\end{equation}
from which the transition temperature
$T_c$ for $N$ atoms is found as the solution of 
\begin{equation}\label{eq:Tc}
N=\frac{\zeta_3(e^{\beta_c V_m},\beta_c V_\sigma,\beta_c V_0)}{\beta_c^3\hbar^3\omega_r^2\omega_z},
\end{equation}
where $\beta_c\equiv1/k_BT_c$. From equations \eeref{eq:N} and \eeref{eq:Tc} we then find the equation of state as
\begin{equation}\label{eq:condFrac}
\frac{N_0}{N}=1-\left(\frac{T}{T_c}\right)^3\frac{\zeta_3(e^{\beta V_m},\beta V_\sigma,\beta V_0)}{\zeta_3(e^{\beta_c V_m},\beta_c V_\sigma,\beta_c V_0)},
\end{equation}
which must be evaluated numerically and describes all the three regimes introduced above: dimple trap, flat trap, and the toroidal trap.

We remark that all ideal gas properties of the system for arbitrary $\sigma_0$ and $V_0$ are not easily obtained from the current formulation. The difficulty arises when considering derivatives of \eeref{zetaDef} with respect to $a$ and $b$, as is required for obtaining the entropy and heat capacity. The function \eeref{zetaDef} is not closed with respect to differentiation, rather leading to a hierarchy of transcendental functions with each successive derivative operation~\cite{Abram}. This problem can be solved by introducing a generalization of \eeref{zetaDef}, and a treatment that includes the double well, toroidal and ellipsoidal cases will be provided elsewhere. For the remainder of this paper we will restrict our attention to the $0\leq V_0$ case and focus primarily on the toroidal regime.

\section{Harmonic scale invariance}
So far we have seen that known results are obtained in the appropriate limits. Our new result is found by using Eq.~(\ref{Vinf}), where we find for the toroidal trap
\begin{equation}\label{FSIR}
{\cal F}=N_0(V_m-\mu)-\frac{\sqrt{2}\pi \sigma_0}{\lambda_{dB}}\frac{\zeta_{7/2}(e^{\beta( \mu-V_m)})}{\beta^3\hbar^2\omega_r\omega_z}.
\end{equation}
What is immediately apparent here is that in this regime all of the ideal gas thermodynamical properties become independent of $V_0$. In particular, the thermodynamics are independent of the toroidal size, and the system enters a \emph{scale invariant regime} defined by
\eqn{\label{SIR}
k_BT\ll V_\sigma\ll V_0.
}
In this regime both of the energy scales $V_0$ and $V_\sigma$ drop out of the problem.
Specifically, for fixed harmonic frequencies $(\omega_r,\omega_z)$, and a beam of fixed width $(\sigma_0)$, the laser power can be increased to generate a toroidal potential with larger perimeter. Nevertheless, all ideal gas properties of the system are invariant under this dilation. The only remnant of the Gaussian beam enters through the length scale $\sigma_0$ appearing in \eref{FSIR}. 

The invariance condition \eeref{SIR} can also be written in terms of appropriate length scales of the trap. We introduce a temperature associated with the optical potential height $V_0=k_B T_0$, and de Broglie wavelength $\lambda_0=\sqrt{2\pi\hbar^2/m k_BT_0}$. Then \eeref{SIR} can be written as
\eqn{\label{lamIneq}
\lambda_0\ll \sqrt{4\pi}\frac{a_r^2}{\sigma_0}\ll\lambda_{dB}.
}
The de Broglie wavelength must be long compared to the other length scales of the system, in the precise sense defined by \eeref{lamIneq}.

We will adopt the more specific term \emph{harmonic scale invariance} for this regime, as we will show below that the condition \eeref{SIR} restricts the system to a regime where the radial trap is well approximated by a quadratic expansion of the potential about the toroidal minimum. In fact, as we shall show in \sref{sec:GenSIC}, the scale invariant property is more general, and the system enters a scale invariant regime whenever $V_\sigma\ll V_0$.
\subsection{Harmonic scale invariance}
A more revealing description of the regime represented by \eeref{FSIR} and \eeref{SIR} is obtained by considering the radial curvature of the trap:
\begin{equation}
\frac{\partial^2 V(\mbf{x})}{\partial r^2}=m\omega_r^2 -\frac{2V_0}{\sigma_0^2}\left(1-\frac{2r^2}{\sigma_0^2}\right)e^{-r^2/\sigma_0^2},
\end{equation}
so that at $r=r_m$, we have
\begin{equation}
\frac{\partial^2 V(\mbf{x})}{\partial r^2}\Bigg{|}_{r=r_m}=\frac{2m\omega_r^2r_m^2}{\sigma_0^2}\equiv m\omega_T^2
\end{equation}
which we have used to define the harmonic radial trapping frequency about the minimum of the toroid
\begin{equation}
\omega_T=\sqrt{2}\frac{\omega_r r_m}{\sigma_0}=\omega_r\sqrt{2\ln{\left(V_0/V_\sigma\right)}}.\label{wT}
\end{equation}

The toroidal trap potential can be expanded about its radial minimum as
\begin{eqnarray}\label{VTdef}
V(\x)&\approx& V_T(\x)\equiv V_m+\frac{m\omega_z^2 z^2}{2}+\frac{m\omega_T^2 (r-r_m)^2}{2}
\end{eqnarray}
with error $O((r-r_m)^3)$.

Using Eq.~(\ref{wT}) for the toroidal frequency, we can write \eref{FSIR} in a more suggestive way:
\begin{equation}\label{Ftoroid}
{\cal F}=N_0(V_m-\mu)-\frac{2\pi r_m}{\lambda_{dB}}\frac{\zeta_{7/2}(e^{\beta( \mu-V_m)})}{\beta^3\hbar^2\omega_T\omega_z},
\end{equation}
where now we see explicitly the scaling with the toroidal perimeter $2\pi r_m$ and the appearance of $\omega_T$ as the physical parameter of the radial degrees of freedom. 

Our treatment thus far has relied on the full semiclassical expression for the free energy for determining the scale invariant regime. A description of the scale invariant regime can also be found by applying the harmonic approximation \eref{VTdef} directly to the free energy \eref{freeE}. In the scale invariant regime given by \eref{SIR}
${\cal G}(\alpha)$ can be approximated by the quadratic expansion of the potential about the minimum \eeref{VTdef}. The integral \eeref{Gdef}, given by
\begin{equation}
{\cal G}(\alpha)=\int_{-\infty}^{\infty} dz\;e^{-\alpha m\omega_z^2z^2/2}\;2\pi\int_0^{\infty} r\;dr\;e^{-\alpha m\omega_r^2r^2/2-\alpha V_0e^{-r^2/2\sigma_0^2}}
\end{equation}
can then be approximated as
\begin{eqnarray}
{\cal G}(\alpha)&\approx&\sqrt{\frac{2\pi}{\alpha m\omega_z^2}}\;2\pi\int_{-r_m}^{\infty} (y+r_m)\;dy\;e^{-\alpha m\omega_T^2y^2/2-\alpha V_m}\nonumber\\
&\approx&\sqrt{\frac{2\pi}{\alpha m\omega_z^2}}\;2\pi r_m\int_{-\infty}^{\infty} \;dy\;e^{-\alpha m\omega_T^2y^2/2-\alpha V_m}\nonumber\\
&=&2\pi r_m\;\frac{2\pi}{\alpha m\omega_z\omega_T}e^{-\alpha V_m}.\label{GdeepT}
\end{eqnarray}
Using this with \eeref{freeE} gives \eeref{Ftoroid} that we previously found using the exact semiclassical free energy in the harmonic scale invariant regime.

Having found the scale invariant grand potential through two different approaches, we now consider the transition temperature for the system, which is given by
\begin{equation}\label{TC1}
T_c=\frac{1}{k_B}\left(\frac{\hbar^2\omega_z\omega_TN}{\zeta(5/2)}\right)^{2/5}\left(\frac{\hbar^2}{\pi m r_m^2}\right)^{1/5}.
\end{equation}
Defining the toroidal kinetic energy 
\eqn{\label{TKE}
\hbar\omega_K=\frac{\hbar^2}{2\pi m r_m^2},
}
and modified geometric mean frequency
\begin{equation}\label{invFreq}
\bar{\omega}^5=\omega_K\omega_T^2\omega_z^2=\frac{\hbar}{\pi m\sigma_0^2}\omega_r^2\omega_z^2,
\end{equation}
the transition temperature becomes
\begin{equation}\label{TcDeepTor}
T_c=\frac{\hbar\bar{\omega}}{k_B}\left(\frac{N}{\zeta(5/2)}\right)^{2/5}\simeq 0.89\frac{\hbar\bar{\omega}}{k_B}N^{2/5}.
\end{equation}
This expression closely resembles that for the three dimensional harmonic trap $k_BT_c=\hbar(\omega_x\omega_y\omega_zN/\zeta(3))^{1/3}$. The scaling with $N^{2/5}$ is caused by the reduction in the number of thermalized degrees of freedom in the system by one, a consequence of toroidal trapping topology. 
The characteristic energy of the system $\hbar \bar{\omega}$, which also determines $T_c$, is seen from \eref{invFreq} to be scale invariant; invariance is caused by the precise dependence of $\omega_T$ on toroidal radius and the fact that this degree of freedom is harmonically bound so that it is doubly weighted in $\bar{\omega}$. In general harmonically bound degrees of freedom have double weight, whereas $\omega_K$ associated with the periodic coordinate is only singly weighted. 

\subsection{Density of states}
The semiclassical density of states is given by 
\eqn{\label{SCdos}
\rho(\epsilon)=\frac{2\pi(2m)^{3/2}}{h^3}\int_{V^*(\epsilon)}d^3\x\;\sqrt{\epsilon-V(\x)},
}
where $V^*(\epsilon)$ is the spatial volume available to a particle with energy $\epsilon$.
Making use of cylindrical symmetry, we obtain
\eqn{\label{symmDos}
\rho(\epsilon)=\frac{1}{\hbar^3}\frac{m}{\omega_z}\int_{R_-}^{R_+} dr\;r[\epsilon-V_{\rm eff}(r)],
}
where $V_{\rm eff}(r)=m\omega_r^2r^2/2+V_0e^{-r^2/\sigma_0^2}-V_m\approx m\omega_T^2(r-r_m)^2/2$ is the shifted radial potential and energy is now expressed relative to $V_m$. The limits of integration are the two solutions of $\epsilon=V_{\rm eff}(R_\pm)$, with $R_-<R_+$.
Making the harmonic approximation to the potential, the semiclassical density of states becomes
\begin{eqnarray}\label{eq:DOStoroid1}
g(\epsilon)&\approx&\frac{1}{\hbar^3}\frac{2mr_m}{\omega_z}\int_0^{\sqrt{2\epsilon/m\omega_T^2}} dy[\epsilon-m\omega_T^2y^2/2],\\\label{eq:DOStoroid2}
&=&\frac{1}{\hbar^3}\frac{2mr_m}{\omega_z}\frac{2\epsilon}{3}\sqrt{\frac{2\epsilon}{m\omega_T^2}}.
\end{eqnarray}
Our analysis of the exact expression of the grand potential has shown that this regime is reached rigorously by taking $V_0\ll V_\sigma $, and $V_\sigma \ll k_BT $ simultaneously. Equation \eeref{eq:DOStoroid2} can be written as
\begin{equation}\label{eq:DOStoroid3}
\lim_{\frac{V_0}{V_\sigma}\to \infty; \frac{V_\sigma}{k_BT}\to \infty}g(\epsilon)\equiv g_T(\epsilon)=\frac{4\epsilon^{3/2}}{3\sqrt{\pi}(\hbar\bar{\omega})^{5/2}},
\end{equation}
giving all thermodynamical properties of the system in the harmonic scale invariant regime. 

\subsection{Thermodynamics}
Using either \eeref{eq:DOStoroid3}, or \eeref{invFreq} and \eeref{Ftoroid}, the grand potential can now be written as
\eqn{\label{GP}
{\cal F}=N_0(V_m-\mu)-\frac{\zeta_{7/2}(e^{\beta( \mu-V_m)})}{\beta(\beta\hbar\bar{\omega})^{5/2}}.
}
For completeness we give the thermodynamic quantities in the harmonic scale invariant regime, both above~($>$) and below~($<$) $T_c$. The total atom number in the system is
\eqn{\label{NtotSIR}
N=N_0+\frac{\zeta_{5/2}(e^{\beta( \mu-V_m)})}{(\beta\hbar\bar{\omega})^{5/2}}.
}
where $N_0>0$ below $T_c$ and the condensate fraction is then given by
\begin{equation}\label{condFracToroid}
\frac{N_0}{N}=1-\left(\frac{T}{T_c}\right)^{5/2},
\end{equation}
again showing the reduction to 5 degrees of freedom.

The entropy $S=-(\partial {\cal F}/{\partial T})_{V,\mu}$ is 
\eqn{
\frac{S_>}{Nk_B}=\frac{7}{2}\frac{\zeta_{7/2}(e^{\beta(\mu-V_m)})}{\zeta_{5/2}(e^{\beta(\mu-V_m)})}-\beta(\mu-V_m),
}
\eqn{\label{ST}
\frac{S_<}{Nk_B}=\frac{7}{2}\frac{\zeta(7/2)}{\zeta(5/2)}\left(\frac{T}{T_c}\right)^{5/2}.
}
The energy $U={\cal F}+TS+\mu N$ is
\eqn{
\frac{U_>}{Nk_BT}=\frac{5}{2}\frac{\zeta_{7/2}(e^{\beta(\mu-V_m)})}{\zeta_{5/2}(e^{\beta(\mu-V_m)})}+\beta V_m,
}
\eqn{
\frac{U_<}{Nk_BT}=\frac{5}{2}\frac{\zeta(7/2)}{\zeta(5/2)}\left(\frac{T}{T_c}\right)^{5/2}+\beta V_m,
}
and the heat capacity $C=(\partial U/\partial T)_{N,V}$ takes the form
\eqn{
\frac{C_>}{Nk_B}=\frac{35}{4}\frac{\zetaa{7/2}{e^{\beta(\mu-V_m)}}}{\zetaa{5/2}{e^{\beta(\mu-V_m)}}}-\frac{25}{4}\frac{\zetaa{5/2}{e^{\beta(\mu-V_m)}}}{\zetaa{3/2}{e^{\beta(\mu-V_m)}}}
}
\eqn{
\frac{C_<}{Nk_B}=\frac{35}{4}\frac{\zeta(7/2)}{\zeta(5/2)}\left(\frac{T}{T_c}\right)^{5/2}.
}
The discontinuity in the heat capacity across the transition $\Delta C(T_c)\equiv C_>(T_c)-C_<(T_c)$ is
\eqn{\label{Cdis}
\frac{\Delta C(T_c)}{Nk_B} = -\frac{25}{4}\frac{\zeta(5/2)}{\zeta(3/2)}\simeq -3.21.
}
\subsection{Generalized volume and pressure and equation of state}
A generalization of thermodynamical formalism to trapped systems has recently been developed and applied by Romero-Rochin {\em et al.}~\cite{RomeroRochin05a,RomeroRochin05b,SandovalFigueroa08a}. This formalism provides a definition of generalized volume and pressure variables for the system, allowing an equation of state to be usefully obtained. While in principle the formalism allows a generalized volume to be defined for any geometry, in general for this system it would have to be determined numerically. To gain some insight into the role of trapping topology in changing effective system volume, in this section we apply this approach to treat the harmonic scale invariant regime of the ideal gas.

Proceeding in a similar manner to \cite{RomeroRochin05a} we see that since $N, U, S$ and ${\cal F}$ are extensive variables and $T$ and $\mu$ are intensive, we may identify the generalized extensive volume in the harmonic scale invariant regime as $\bar{\omega}^{-5/2}$:
\eqn{
{\cal V}\equiv(\omega_K\omega_T^2\omega_z^2)^{-1/2}.
}
That it takes this form is not so surprising, since at a given temperature the gas will be mainly confined to a volume of order (toroidal circumference)$\times$(cross sectional area) $\sim 2\pi r_m(k_BT/m\omega_z^2)^{1/2}(k_BT/m\omega_T^2)^{1/2}\propto \bar{\omega}^{-5/2}$. This should be compared to the volume for a harmonically trapped gas which is of order $(k_BT/m\omega^2)^{3/2}\propto \omega^{-3}$, where $\omega=(\omega_x\omega_y\omega_z)^{1/3}$ is the usual geometric mean~\cite{RomeroRochin05a}. We can now identify an important change in the system relative to the purely harmonic case: the system has effectively two thermally determined length scales parameterizing the cross section, and one purely geometric length scale that is under direct experimental control, the toroidal circumference.

The conjugate generalized intensive pressure ${\cal P}=-(\partial {\cal F}/\partial {\cal V})_{N,T}$ is then given by
\eqn{\label{Pgen}
{\cal P}=-\frac{\zeta_{7/2}(e^{\beta( \mu-V_m)})}{\beta(\beta\hbar)^{5/2}}.
}
Above $T_c$ this gives the equation of state ${\cal PV}=-{\cal F}$ as required~\cite{RomeroRochin05a}.

It is important to see that the thermodynamic limit is well defined in the scale invariant regime. The appropriate limit is $N\to \infty$, ${\cal V}\to \infty$, with $N/{\cal V}=N\bar{\omega}^{5/2}\to$ constant. Since ${\cal V}\propto \sigma_0/\omega_r\omega_z$ we see that thermodynamic limit requires 
\eqn{\label{Vrat}
\frac{N\omega_r\omega_z}{\sigma_0}\to {\rm constant},
}
corresponding to either relaxing the harmonic trapping frequencies, or increasing the Gaussian beam width $\sigma_0$ appropriately. For the latter case, if the system is to remain in the scale invariant regime ($V_\sigma\ll V_0$) we must also impose the further condition that $V_0\to \infty$ at least as fast as $\sigma_0^2$ to obtain a consistent thermodynamic limit in this regime.
\section{General scale invariance}\label{sec:GenSIC}
\begin{figure}[tbp]
\includegraphics[width=.9\columnwidth]{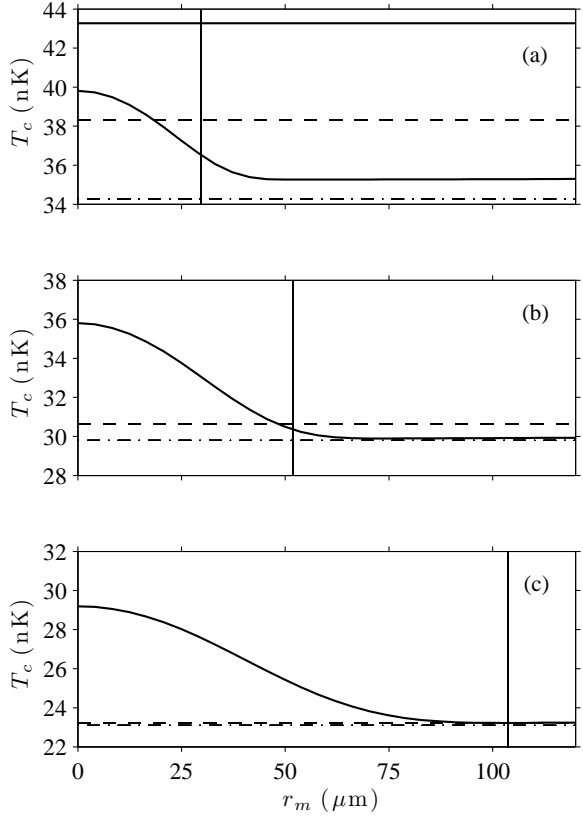}
\caption{Numerical $T_c$ for $^{87}$Rb atoms held in a cylindrically symmetric harmonic trap for a range of Gaussian potentials parametrized by the location of the trap minimum $r_m$. $T_c$ (solid curves) is calculated for (a) $\sigma_0=14.1\mu$m, (b) $\sigma_0=24.7\mu$m, and (c) $\sigma_0=49.5\mu$m, by solving \eeref{eq:Tc}. The solid horizontal line in (a) shows $T_c=43.3{\rm nK}$ for the purely harmonic trap ($V_0=0$). Dashed lines give the harmonic scale invariant $T_c^0$ \eeref{TcDeepTor}, and dash-dotted lines show the correction $T_c^0+\delta T_c^1$ given by \eeref{firstOrdCor} . The ratio $k_BT_c(V_0\to\infty)/V_\sigma$ is 3.5, 0.95, and 0.19 for (a), (b), and (c) respectively, and the vertical lines show $V_0=3V_\sigma$. Other parameters are $N=10^6$ atoms and $(\omega_z,\omega_r)=2\pi (15.3,7.8)$ Hz.}
\label{fig2}
\end{figure}

In this section we show that it is possible to further relax the condition \eeref{SIR} for scale invariance. Our numerical investigations show that a scale invariant regime is always reached provided $V_\sigma\ll V_0$ (see  \fref{fig2}). Thus there are scale invariant regimes parameterized by the ratio $V_\sigma/k_B T\sim 1$, corresponding to $\lambda_{dB}\sim  a_r^2/\sigma_0$.

However, the properties of the system are not well described by the harmonic approximation to the potential \eeref{VTdef} unless $k_B T\ll V_\sigma$ also holds. We now seek a general asymptotic expansion of the grand potential which does not require this condition to hold. 

We make use of the representation of $\gamma(a,x)$ in terms of the confluent hypergeometric function $M(a,b,z)$~\cite{Abram} $\gamma(a,x)=a^{-1}x^a e^{-x}M(1,1+a,x)$,
and the asymptotic expansion of $M(a,b,z)$ for large real positive $z$ and fixed $a, b$. To leading order in powers of $z^{-1}$ we easily obtain
\eqn{\label{gammaExp}
\gamma(a,x)=\left(a^{-1}\Gamma(a+1)-z^{a-1}e^{-z}\right)(1+O(z^{-1})).
}
At leading order in $(\beta V_0)^{-1}$ we then find 
\eqn{
\zeta_\nu(e^{\beta\mu},\beta V_\sigma,\beta V_0)&\simeq& \sum_{k=1}^\infty \frac{e^{k\beta\mu}}{k^\nu}\frac{\Gamma(k\beta V_\sigma +1)}{(k\beta V_0)^{k\beta V_\sigma}}\nonumber\\
&&-\frac{V_\sigma}{V_0}\zeta_\nu(e^{\beta(\mu-V_0)}).
}
As $V_0/V_\sigma \to \infty$, $\zeta_\nu(e^{\beta(\mu-V_0)})\to e^{\beta(\mu-V_0)}$ and the second term vanishes. We note that, for the vast majority of terms in the summation, the inequality $1\ll k\beta V_\sigma$ will hold, even if $\beta V_\sigma\sim 1$. Introducing Stirling's expansion for $\Gamma(k\beta V_\sigma+1)$ and making use of the identity $e^{-\beta V_m}=e^{-\beta V_\sigma}(V_\sigma / V_0)^{\beta V_\sigma}$ we finally obtain the expansion
\eqn{\label{genSI}
\lim_{\frac{V_0}{V_\sigma}\to \infty}\frac{\zeta_\nu(e^{\beta\mu},\beta V_\sigma,\beta V_0)}{\sqrt{2\pi\beta V_\sigma}}&=&\Bigg[\zeta_{\nu-1/2}(e^{\beta(\mu-V_m)})\nonumber\\
&&+\frac{\zeta_{\nu+1/2}(e^{\beta(\mu-V_m)})}{12\beta V_\sigma}\nonumber\\
&&+\frac{\zeta_{\nu+3/2}(e^{\beta(\mu-V_m)})}{288(\beta V_\sigma)^2}-\dots\Bigg],
}
where the numerical coefficients of the asymptotic expansion are the coefficients of the Stirling expansion for $\Gamma(z)$. This can be used to give a more general expression for the free energy than the asymptotic form \eeref{GP} which arises from the leading term in \eeref{genSI}.

In the regime where $V_\sigma\ll V_0$ the grand free energy can now be written as
\eqn{\label{SIF}
{\cal F}&=&N_0(V_m-\mu)-\frac{1}{\beta(\beta\hbar\bar{\omega})^{5/2}}\Bigg[\zeta_{7/2}(e^{\beta(\mu-V_m)})\nonumber\\
&&+\frac{\zeta_{9/2}(e^{\beta(\mu-V_m)})}{12\beta V_\sigma}\nonumber\\
&&+\frac{\zeta_{11/2}(e^{\beta(\mu-V_m)})}{288(\beta V_\sigma)^2}-\dots\Bigg].
}
This expression for the free energy is our main result and provides an asymptotically exact representation of all thermodynamics of the system in the scale invariant regime. The energy scale $V_0$ only appears through the shift of the potential minimum $V_m$. Ignoring this trivial shift of energy, we now have 
\eqn{\label{genSICF}
\lim_{\frac{V_0}{V_\sigma}\to \infty}\frac{\partial {\cal F}}{\partial r_m}\equiv 0,
}
and all ideal gas thermodynamics of the system are scale invariant. In practice this regime is reached rather quickly. In this example shown in \fref{fig2} the onset of scale invariance occurs around $V_0\sim 3V_\sigma$.

\subsection{Non-harmonic corrections to $T_c$}
As an application of the generalized expansion \eeref{SIF} we can now find a more accurate expression for $T_c$. We easily find the following asymptotic expansion for $T_c$:
\eqn{\label{Tcmod}
\frac{T_c}{T_c^0}&=&\left[1+\frac{k_B T_c\zeta(7/2)}{12V_\sigma\zeta(5/2)}+\frac{(k_B T_c)^2\zeta(9/2)}{288V_\sigma^2\zeta(5/2)}-\dots\right]^{-2/5}
}
where the atom number has been eliminated in terms of the harmonic scale invariant transition temperature \eeref{TcDeepTor} denoted by $T_c^0$. We now expand $T_c=T_c^0+\delta T_c^1+\delta T_c^2+\dots$, which we treat formally as a perturbation expansion in powers of $k_B T_c^0/V_\sigma$. Solving for the first order correction, we find
\eqn{\label{firstOrdCor}
\frac{\delta T_c^1}{T_c^0}=-\frac{k_BT_c^0}{30V_\sigma}\frac{\zeta(7/2)}{\zeta(5/2)}\simeq -1.44\times 10^{-2}\frac{k_BT_c^0}{V_\sigma}.
}
In \fref{fig2} we show the numerical solution of \eeref{eq:Tc} and compare with the harmonic scale invariant form \eeref{TcDeepTor} and the first correction for non-harmonic behavior \eeref{firstOrdCor}. The full numerical solution reaches a scale invariant regime for $V_\sigma \ll V_0$ and approaches the harmonic behavior for $k_B T_c\ll V_\sigma$. The rapid decay of the Stirling expansion coefficients renders the first correction given by \eeref{firstOrdCor} adequate even when $k_B T_c^0\sim 3V_\sigma$, as can be seen from \fref{fig2} (a) where $T_c^0+\delta T_c^1$ approaches $T_c$ to within $3\%$. 
\section{Non-ideal corrections to $T_c$}
The ideal gas behavior reveals a scale invariant regime for the toroidal trap. While this gives a strong indication of what to expect in experiments, it is important to determine how additional effects will change this picture. We now consider the two most significant effects on the transition temperature: finite size corrections, and interactions. Our main focus here is on the modifications these effects will make to the scale invariance of quantum degeneracy.

\subsection{Mean field interaction shift in $T_c$}
Recent experimental~\cite{Gerbier04a} and theoretical~\cite{Davis2006a} studies of the harmonically trapped Bose gas have established that the value for $T_c$ is well described by the combination of ideal gas theory and the first order mean field interaction shift~\cite{Giorgini96a}. Within experimental error bars, the mean field interaction shift is the dominant effect and critical fluctuations appear to be entirely negligible. 
Since the system we consider is in the semiclassical regime we evaluate the first order interaction shift due to s-wave collisions from the expression derived by Giorgini \etal~\cite{Giorgini96a}, which is 
\begin{equation}\label{dTint}
\frac{\delta T_c^{int}}{T_c^0}=-\frac{2U_0}{T_c^0}\frac{\int d^3\x\; \partial n_{th}^0/\partial \mu\;[n_{th}^0(\x=0)-n_{th}^0(\x)]}{\int d^3\x\; \partial n_{th}^0/\partial T}
\end{equation}
where $n_{th}^0(\x)$ is the non-interacting semiclassical thermal cloud density
\begin{equation}\label{nT0}
n_{th}^0(\mbf{x})=\frac{1}{\lambda_{dB}^3}\sum_{k=1}^\infty \frac{e^{-\beta k(\mu-V(\mbf{x}))}}{k^{3/2}},
\end{equation}
and $U_0=4\pi\hbar^2 a/m$ gives the interaction strength in terms of the s-wave scattering length $a$.
As before, in the harmonic scale invariant regime where \eeref{SIR} holds, we can carry out the harmonic approximation for the potential.
A straightforward calculation similar to that of Ref.~\cite{Giorgini96a} gives
\begin{equation}\label{dTintExp}
\frac{\delta T_c^{int}}{T_c^0}=-\frac{aN^{1/5}}{\bar{a}}\left(\frac{8}{5\sqrt{2}}\frac{\zeta(3/2)^2(1-G)}{\pi^{1/2}\zeta(5/2)^{6/5}}\right),
\end{equation}
where $\bar{a}=\sqrt{\hbar/m\bar{\omega}}$ is the modified geometric mean of the toroid length scales defined as $\bar{a}^5=a_Ka_T^2a_z^2$, with $a_j=\sqrt{\hbar/m\omega_j}$, and
\begin{equation}
G=\frac{1}{\zeta(3/2)^2}\sum_{j=1,k=1}^\infty\frac{1}{j^{3/2}k^{1/2}(j+k)}\equiv \frac{S}{\zeta(3/2)^2}
\end{equation}
We can obtain $G$ by writing
\eqn{\label{Ssolve}
S&=&\sum_{j=1,k=1}^\infty\left(1-\frac{j}{j+k}\right)\frac{1}{j^{3/2}k^{3/2}}=\zeta(3/2)^2-S
}
to give $G=1/2$.
Evaluating the numerical factors in \eeref{dTintExp} gives
\begin{equation}\label{dTintNum}
\frac{\delta T_c^{int}}{T_c^0}\simeq-1.53\frac{aN^{1/5}}{\bar{a}}.
\end{equation}
This expression has the same structure as the well known result for the shift in $T_c$ in the harmonic trap~\cite{Giorgini96a}. A different characteristic length scale arises here and the dependence is now on $N^{1/5}$ rather than the $N^{1/6}$ behavior in the three dimensional harmonic trap.
As noted in Ref.~\cite{Giorgini96a}, the first order interaction shift depends on the geometric mean frequency of the trap. Here we see the interaction shift is invariant with the size of the toroid, depending on system size only through the modified geometric mean length scale $\bar{a}$. 

\subsection{Finite size effect on $T_c$}
The effect of finite particle number is calculated to first order by shifting the ground state chemical potential up to the quantum mechanical ground state of the potential \eeref{VTdef}:
\begin{equation}
\delta\mu=\frac{\hbar}{2}(\omega_T+\omega_z),
\end{equation}
which shifts $T_c$ according to
\begin{equation}\label{dTfs}
\frac{\delta T_c^{fs}}{T_c^0}=\frac{1}{T_c^0}\frac{\partial T_c^0}{\partial \mu}\Bigg|_{N}\delta\mu=-\frac{\zeta(3/2)}{5\zeta(5/2)}\left(\frac{\zeta(5/2)}{N}\right)^{2/5}\frac{\omega_T+\omega_z}{\bar{\omega}}
\end{equation}
Evaluating the numerical factors gives
\eqn{
\frac{\delta T_c^{fs}}{T_c^0}\simeq-0.44\frac{\omega_T+\omega_z}{\bar{\omega}}N^{-2/5}
}
As $\omega_T\propto r_m$ while $\bar{\omega}$ is invariant, finite size effects can become a significant correction for large $r_m$. However, the scaling with $N^{-2/5}$ will strongly suppress this effect for large $N$.

For the parameters used in \fref{fig2} (c), the shifts have the values: $\delta T_c^{int}/T_c^0\sim -2\times10^{-5}$  (for $100\mu m<r_m$), and $-2\times 10^{-2}<\delta T_c^{fs}/T_c^0< -6\times 10^{-2}$ (for $100\mu m <r_m<500 \mu m$). In contrast to the harmonically trapped gas, for these parameters the finite size correction is the dominant shift in the toroidal trap. However, the finite size shift vanishes in the thermodynamic limit and for significant atom number (here $N=10^6$) there is a wide range of toroidal radii where scale invariance of $T_c$ holds to within a few percent.

\section{Reaching scale invariance}
There are at least three means to reach the scale invariant regime with degenerate Bose gases. Firstly, a non-condensed gas may be evaporatively cooled below $T_c$ into a toroidal trap as recently demonstrated experimentally~\cite{Weiler08a}. A second method is to adiabatically load from the harmonic trap into the toroid by ramping up the optical potential which we further investigate below. Lastly, a gas in a harmonic trap with a small optical potential may be rotated which will push it into the scale invariant regime. 

The physics of the rotating case is rather simple: an ideal Bose gas in equilibrium in the trap \eeref{Vdef} in a frame rotating around the $z$-axis with frequency $\Omega<\omega_r$ has thermodynamic properties of a system in the lab frame with effective radial trapping frequency $\omega_\perp=\sqrt{\omega_r^2-\Omega^2}$~\cite{Bradley08a}. Thus all the properties of the system follow from the lab frame analysis presented above, after replacing $V_\sigma$ with an effective rotating frame energy $V_\sigma^\perp=m\omega_\perp^2\sigma_0^2/2=V_\sigma(1-\Omega^2/\omega_r^2)$. In the toroidal regime (where $V_\sigma^\perp< V_0$) the radius to the trap minimum $r_m^\perp$ is given by 
\eqn{\label{rperp}
r_m^\perp=\sigma_0\left[\ln{(V_0/V_\sigma)}+\ln{\left(\frac{\omega_r^2}{\omega_r^2-\Omega^2}\right)}\right]^{1/2}.
}
The essential condition for scale invariance becomes 
\eqn{
V_\sigma (1-\Omega^2/\omega_r^2) \ll V_0
}
for the rotating gas, and the harmonic approximation will be valid when $k_BT\ll V_\sigma (1-\Omega^2/\omega_r^2)$ also holds.
The frequency \eeref{invFreq} determining harmonic scale invariant properties is modified to 
\eqn{
\bar{\omega}_\perp^5=\frac{\hbar}{\pi m\sigma_0^2}\omega_r^2\omega_z^2(1-\Omega^2/\omega_r^2)
}
by the rotation. We conclude that, for a gas in rotating equilibrium, scale invariance with respect $V_0$ is preserved, but with appropriately modified characteristic energy $\hbar\bar{\omega}_\perp$ determining the density of states. However, since $\bar{\omega}_\perp$ depends on $\Omega$, changes in the toroidal radius \eeref{rperp} caused by rotation do not have the scale invariant property.
\begin{figure}[tbp]
\includegraphics[width=\columnwidth]{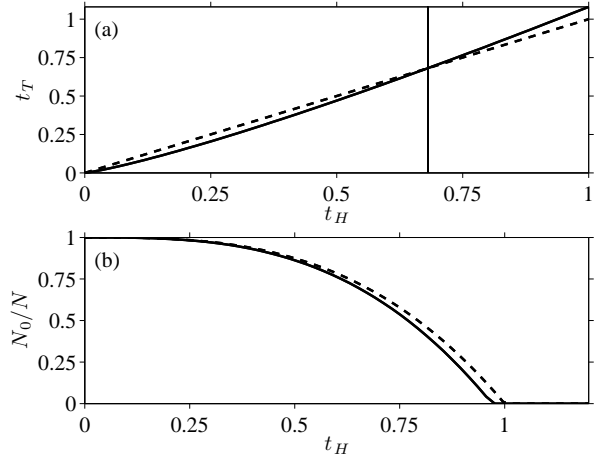}
\caption{Adiabatic loading from a harmonic trap into the scale invariant regime. (a) Reduced temperature in the toroidal trap (solid line) after adiabatically ramping up the optical potential for a system with reduced temperature $t_H$ in the harmonic trap. The line $t_T=t_H$ (dashed line) is shown for comparison. The point $t_H=t_H^*\simeq 0.67$, from \eeref{tHdef}, is shown by the vertical line. (b) Condensate fraction in the toroidal trap (solid line) after adiabatic loading from initial reduced temperature $t_H$. The condensate fraction for the harmonic trap is also shown (dashed line).}
\label{fig3}
\end{figure}
\subsection{Adiabatic loading}
We consider adiabatically loading a degenerate Bose gas from the harmonic trap into the toroidal trap. In order to maintain adiabaticity the timescale of loading $T_L$ should greatly exceed the slowest timescale of the system, i.e: ${\rm min}(2\pi/\omega_j)\ll T_L$. Under these conditions entropy will be conserved during loading. Equating the entropy in the harmonic trap for the regime $T<T_c$
\begin{equation}\label{SH}
\frac{S_<^{H}}{Nk_B}=\frac{4\zeta(4)}{\zeta(3)}\left(\frac{T}{T_c}\right)^3\equiv \frac{4\zeta(4)}{\zeta(3)}t_H^3,
\end{equation}
with the scale invariant toroidal entropy of \eref{ST} 
\begin{equation}
\frac{S_<^{T}}{Nk_B}=\frac{7\zeta(7/2)}{\zeta(5/2)}t_T^{5/2},
\end{equation}
we find
\begin{equation}
t_T(t_H)=\left(\frac{8\zeta(4)\zeta(5/2)}{7\zeta(3)\zeta(7/2)}\right)^{2/5}t_H^{6/5}
\end{equation}
for the reduced temperature in the scale invariant toroid after adiabatic loading from the harmonic trap. From this expression we can find $t_H^*$ defined by $t_T(t_H^*)\equiv t_H^*$, ie: where adiabatic loading has no effect on the reduced temperature in the toroid. This is given by
\eqn{\label{tHdef}
t_H^*=\left(\frac{7\zeta(3)\zeta(7/2)}{8\zeta(4)\zeta(5/2)}\right)^{2}\simeq 0.67.
}
Evaluating the numerical factors gives
\begin{equation}\label{eq:TH}
t_T=t_H^{6/5}/(t_H^*)^{1/5}\simeq 1.08 t_H^{6/5},
\end{equation}
which is shown in \fref{fig3} (a). We see that to a good approximation $t_T\approx t_H$ for the region $t_H\leq 1$. Substituting \eref{eq:TH} into \eref{condFracToroid}, gives the expression
\begin{equation}
\left(\frac{N_0}{N}\right)=1-t_H^{3}/(t_H^*)^{1/2}\simeq 1-1.23t_H^{3}
\end{equation}
for the condensate fraction in the toroid after loading. This expression is shown in \fref{fig3} (b), where we see that the condensate fraction is well preserved during loading from the purely harmonic trap to the harmonic scale invariant toroidal trap.

\section{Conclusions}
We have provided a numerical and analytical treatment of the Bose gas trapped in a harmonic-Gaussian trap with toroidal topology. 
We have identified a regime where the harmonic-Gaussian toroidal trap has properties that are independent of the radius to the toroidal minimum. In particular, the transition temperature to Bose-Einstein condensation is scale invariant, and the toroid radius can then be increased without altering quantum degeneracy. 

\emph{Harmonic scale invariance.---}In the regime where $k_B T_c\ll V_\sigma= m\omega_r^2\sigma_0^2/2$, the radial trap is well approximated by its quadratic expansion about the minimum. This regime affords an analytical description and we have identified the invariant generalized extensive volume ${\cal V}=(\omega_K\omega_z^2\omega_T^2)^{-1/2}$ which determines the thermodynamics of the system. We have also shown that the first order mean field interaction shift to $T_c$ is scale invariant in this regime. The main limitation on the invariance of $T_c$ is the finite size shift that vanishes in the thermodynamic limit.

\emph{General scale invariance.---}Relaxing the condition $k_B T_c\ll V_\sigma$, we find that the ideal gas always enters a scale invariant regime when $V_\sigma\ll V_0$, for which the grand potential becomes independent of the toroidal radius: $\partial {\cal F}/\partial r_m=0$. In considering corrections to the harmonic approximation to the potential we have evaluated the first order perturbation of $T_c$ in powers of $k_BT_c/V_\sigma$, and find it provides an accurate approximation even in the regime where $V_\sigma\sim k_B T_c$. In practice scale invariance is reached quite rapidly with increasing $V_0$, and the onset occurrs at about $V_0\sim 3V_\sigma$ for the specific system treated here. 

\emph{Reaching scale invariance.---}We have considered adiabatic loading and applying rotation as ways to reach the scale invariant regime. Adiabatic loading appears to preserve the condensate fraction of the ideal gas quite well, while rotation enhances the height of the Gaussian relative to the rotating frame harmonic trap, thus deepening the toroidal potential. Another promising method is to evaporatively cool directly into the toroidal trap. The existence of a scale invariant regime shows that rather than always decreasing with toroidal radius, $T_c$ for such a system reaches a well defined plateau. A system with these properties may be promising for creating and loading a large toroidal trap with a persistent current while maintaining quantum degeneracy.
Future work will focus on more general thermodynamical quantities, the role of interactions, and extensions to trapped Fermi gases.

\subsection*{Acknowledgements}
I am grateful to P. B. Blakie for a critical reading of this manuscript, and to
B. P. Anderson, D. Baillie, M. J. Davis, I. B. Kinski, M. K. Olsen, and E. van Oojen for useful discussions.
It is a pleasure to thank the Queensland node of the ARC Centre of Excellence for Quantum-Atom Optics where this work began.
This work was supported by the New 
Zealand Foundation for Research Science and Technology under contract UOOX0801.

\bibliographystyle{prsty}

\begin{thebibliography}{10}

\bibitem{Anglin1999}
J.~R. Anglin and W.~H. Zurek, Phys. Rev. Lett. {\bf 83},  1707  (1999).

\bibitem{Dziarmaga2008a}
J. Dziarmaga, J. Meisner, and W.~H. Zurek, Phys. Rev. Lett. {\bf 101},  115701
  (2008).

\bibitem{Kibble1976}
T.~W.~B. Kibble, J. Phys. A: Math. Gen. {\bf 9},  1387  (1976).

\bibitem{Zurek1985}
W.~H. Zurek, Nature {\bf 317},  505  (1985).

\bibitem{Ryu07a}
C. Ryu, M.~F. Andersen, P. Clad\'{e}, V. Natarajan, K. Helmerson, and W.~D.
  Phillips, Phys. Rev. Lett. {\bf 99},  260401  (2007).

\bibitem{Weiler08a}
C.~N. Weiler, T.~W. Neely, D.~R. Scherer, A.~S. Bradley, M.~J. Davis, and B.~P.
  Anderson, Nature {\bf 455},  948  (2008).

\bibitem{Sauer01a}
J. A. Sauer, M. D. Barrett, and M. S. Chapman, Phys. Rev. Lett. {\bf 87},  270401
  (2001).

\bibitem{Wright00a}
E.~M. Wright, J. Arlt, and K. Dholakia, Phys. Rev. A {\bf 63},  013608  (2000).

\bibitem{Arnold04a}
A.~S. Arnold, J. Phys. B: At. Mol. Opt. Phys. {\bf 37},  L29  (2004).

\bibitem{Gustavson00a}
T. Gustavson, A. Landragin, and M. Kasevich, Classical Quant Grav {\bf 17},
  2385  (2000).

\bibitem{Arnold06a}
A.~S. Arnold, C.~S. Garvie, and E. Riis, Phys. Rev. A {\bf 73},  041606(R)
  (2006).

\bibitem{Gupta2005}
S. Gupta, K.~W. Murch, K.~L. Moore, T.~P. Purdy, and D.~M. Stamper-Kurn, Phys.
  Rev. Lett. {\bf 95},  143201  (2005).

\bibitem{Heathcote08a}
W.~H. Heathcote, E. Nugent, B.~T. Sheard, and C.~J. Foot, N. J. Phys. {\bf 10},
   043012  (2006).

\bibitem{Schnelle08a}
S.~K. Schnelle, E.~D. van Ooijen, M.~J. Davis, N.~R. Heckenberg, and H.
  Rubinsztein-Dunlop, Optics Express {\bf 16},  1405  (2008).

\bibitem{Petrosyan99a}
K. G. Petrosyan and L. You, Phys. Rev. A {\bf 59},  639  (1999).

\bibitem{Benakli99a}
M. Benakli, S. Raghavan, A. Smerzi, S. Fantoni, and S.~R. Shenoy, Europhys.
  Lett. {\bf 46},  275  (1999).

\bibitem{Ogren07}
M. \"{O}gren and G.~M. Kavoulakis, J. Low Temp. Phys. {\bf 154},  149  (2007).

\bibitem{Ogren09}
M. \"{O}gren and G.~M. Kavoulakis, J. Low Temp. Phys. {\bf 154},  30  (2009).

\bibitem{Javanainen98a}
J. Javanainen, S. M. Paik, and S. M. Yoo, Phys. Rev. A {\bf 58},  580  (1998).

\bibitem{Salasnich99a}
L. Salasnich, A. Parola, and L. Reatto, Phys. Rev. A {\bf 59},  2990  (1999).

\bibitem{Jackson06a}
A. D. Jackson and G.~M. Kavoulakis, Phys. Rev. A {\bf 74},  065601  (2006).

\bibitem{Nugent03a}
E. Nugent, D. McPeake, and J.~F. McCann, Phys. Rev. A {\bf 68},  063606
  (2003).

\bibitem{Anderson03a}
B.~P. Anderson, K. Dholakia, and E.~M. Wright, Phys. Rev. A {\bf 67},  033601
  (2003).

\bibitem{Schulte02a}
T. Schulte, L. Santos, A. Sanpera, and M. Lewenstein, Phys. Rev. A {\bf 66},
   033602  (2002).

\bibitem{Brand01a}
J. Brand and W. Reinhardt, J Phys B-At Mol Opt {\bf 34},  L113  (2001).

\bibitem{Jain07a}
P. Jain, A.~S. Bradley, and C.~W. Gardiner, Phys. Rev. A {\bf 76},  023617  (2007).

\bibitem{Modugno06a}
M. Modugno, C. Tozzo, and F. Dalfovo, Phys. Rev. A {\bf 74},  061601(R)  (2006).

\bibitem{Bhattacherjee04a}
A.~B. Bhattacherjee, E. Courtade, and E. Arimondo, J Phys B-At Mol Opt {\bf
  37},  4397  (2004).

\bibitem{Kavoulakis04a}
G.~M. Kavoulakis, Phys. Rev. A {\bf 69},  023613  (2004).

\bibitem{Cozzini05a}
M. Cozzini, A.~L. Fetter, B. Jackson, and S. Stringari, Phys. Rev. Lett. {\bf
  94},  100402  (2005).

\bibitem{Cozzini06a}
M. Cozzini, B. Jackson, and S. Stringari, Phys. Rev. A {\bf 73},  013603  (2006).

\bibitem{Kling08a}
S. Kling and A. Pelster, Phys. Rev. A {\bf 76},  023609  (2007).

\bibitem{Bretin04a}
V. Bretin, S. Stock, Y. Seurin, and J. Dalibard, Phys. Rev. Lett. {\bf 92},
  050403  (2004).

\bibitem{Blakie07a}
P.~B. Blakie and W. X. Wang, Phys. Rev. A {\bf 76},  053620  (2007).

\bibitem{Bagnato87a}
V. Bagnato, D.~E. Pritchard, and D. Kleppner, Phys. Rev. A {\bf 35},  4354
  (1987).

\bibitem{RomeroRochin05a}
V. Romero-Rochin, Phys. Rev. Lett. {\bf 94},  130601  (2005).

\bibitem{Grimm01a}
R. Grimm, M. Weidemuller, and Y. Ovchinnikov, {\em {Optical dipole traps for
  neutral atoms}}, Vol.~{42} of {\em Advances in Atomic Molecular and Optical
  Physics} (Academic Press, San Diego, USA, {2000}), pp.\ {95--170}.

\bibitem{Abram}
M. Abramowitz and I.~A. Stegun, {\em Handbook of Mathematical Functions}, ninth
  ed. (Dover, New York, 1972).

\bibitem{RomeroRochin05b}
V. Romero-Rochin and V. Bagnato, Braz. J. Phys. {\bf 35},  607  (2005).

\bibitem{SandovalFigueroa08a}
N. Sandoval-Figueroa and V. Romero-Rochin, Phys. Rev. E {\bf 78},  061129
  (2008).

\bibitem{Gerbier04a}
F. Gerbier, J.~H. Thywissen, S. Richard, M. Hugbart, P. Bouyer, and A. Aspect,
  Phys. Rev. Lett. {\bf 92},  030405  (2004).

\bibitem{Davis2006a}
M.~J. Davis and P.~B. Blakie, Phys. Rev. Lett. {\bf 96},  060404  (2006).

\bibitem{Giorgini96a}
S. Giorgini, L.~P. Pitaevskii, and S. Stringari, Phys. Rev. A {\bf 54},  R4633
  (1996).

\bibitem{Bradley08a}
A.~S. Bradley, C.~W. Gardiner, and M.~J. Davis, Phys. Rev. A {\bf 77},  033616
  (2008).

\end{thebibliography}

\end{document}